\documentclass{ifacconf}

\usepackage{graphicx}% include this line if your document contains figures
\usepackage{natbib} % required for bibliography
\usepackage{amsmath}
\usepackage{amssymb}
\usepackage{algorithm}
\usepackage{algpseudocode}
\usepackage{booktabs}
\usepackage{multirow}
\usepackage{xcolor}
\usepackage{soul}
\usepackage{rotating} 
\usepackage{url}

\begin{document}
\begin{frontmatter}

%{\color{red} \textbf{© 2023 the authors. This work has been accepted to IFAC for publication under a Creative Commons Licence CC-BY-NC-ND}}

{\color{red} \textbf{© 2024 the authors. This work has been accepted to BMS-IFAC
for publication under a Creative Commons Licence
CC-BY-NC-ND.}}

\title{Attention Networks for Personalized Mealtime Insulin Dosing in People with Type 1 Diabetes \thanksref{footnoteinfo}} 

\thanks[footnoteinfo]{Supported by American Diabetes Association grant \#4‐22‐PDFPM‐16.}

\author[First]{Anas El Fathi} 
\author[Second]{Elliott Pryor} 
\author[Third]{Marc D. Breton} 

\address[First]{University of Virginia, Center for Diabetes Technology, Charlottesville, VA, USA (e-mail: fwt9vd@virginia.edu).}
\address[Second]{University of Virginia, Center for Diabetes Technology, Charlottesville, VA, USA (e-mail: hyy8sc@virginia.edu)}
\address[Third]{University of Virginia, Center for Diabetes Technology, Charlottesville, VA, USA (e-mail: mb6nt@virginia.edu)}

\begin{abstract} % Abstract of not more than 250 words.
Calculating mealtime insulin doses poses a significant challenge for individuals with Type 1 Diabetes (T1D). Doses should perfectly compensate for expected post-meal glucose excursions, requiring a profound understanding of the individual's insulin sensitivity and the meal macronutrients'. Usually, people rely on intuition and experience to develop this understanding. In this work, we demonstrate how a reinforcement learning agent, employing a self-attention encoder network, can effectively mimic and enhance this intuitive process. Trained on 80 virtual subjects from the FDA-approved UVA/Padova T1D adult cohort and tested on twenty, self-attention demonstrates superior performance compared to other network architectures. Results reveal a significant reduction in glycemic risk, from 16.5 to 9.6 in scenarios using sensor-augmented pump and from 9.1 to 6.7 in scenarios using automated insulin delivery. This new paradigm bypasses conventional therapy parameters, offering the potential to simplify treatment and promising improved quality of life and glycemic outcomes for people with T1D.

\end{abstract}

\begin{keyword}
Type 1 Diabetes, Reinforcement Learning, Self-Attention, Postprandial Glucose.
\end{keyword}

\end{frontmatter}
%===============================================================================

\section{Introduction}

\subsection{Background and Motivation}

Type 1 diabetes (T1D) is an autoimmune disease treated by exogenous insulin to maintain blood glucose levels at recommended targets. Sustained elevated glucose levels (hyperglycemia) lead to long-term micro-/macro-vascular complications. 
Fear of low glucose levels (hypoglycemia) and its acute complications are major limiting factors to achieving glucose targets. 
T1D treatment involves intensive insulin therapy, in which basal and bolus insulin is administered, and glucose levels are regularly monitored. 
Basal insulin aims to maintain glucose levels constant during fasting and overnight. 
Bolus insulin is taken at mealtimes to compensate for the significant glucose increase due to meal carbohydrates. 
Insulin therapy can be implemented as ``Multiple Daily Injections" (MDI) of insulin utilizing insulin syringes or pens.
However, the gold standard of insulin therapy is ``Automated Insulin Delivery" (AID) systems employing an insulin pump and a continuous glucose monitoring (CGM) system. 
AID systems dynamically adjust the rate of insulin delivery based on real-time CGM.

Regardless of the treatment modality, individuals with T1D are required to meticulously calculate their insulin doses for meals \citep{el2018artificial}.

\subsection{Traditional Bolus Calculators for T1D} \label{sec:bolusCalc}
Since carbohydrates are the primary determinant of the post-meal response, people with T1D are encouraged to perform carbohydrate (CHO) counting to calculate a mealtime insulin bolus \citep{sheard2004dietary}. In practice, proper CHO counting requires: a rigorous education, good numeracy skills, is prone to human errors, and is associated with increased disease management burden \citep{fortin2017practices, mannucci2005eating, brazeau2013carbohydrate}. In addition, CHO counting involves the use of individualized parameters such as carbohydrate ratios (CR), a ratio indicating the amount of carbohydrate covered by a unit of insulin, the correction factor (CF), a factor indicating the drop in glucose levels caused by a unit of insulin, and the glucose target. These therapy parameters require continuous adaptation and optimization. 
Equation \ref{eq:standardbolus} shows an example of the most common formula used to calculate the insulin dose at mealtime $B$ using these parameters, where IOB indicates the remaining active insulin on board.

%This complex calculation encompasses a bolus dose for the carbohydrate content of the meal, a corrective dose to mitigate any elevated glucose levels, and a tailored adjustment for insulin-on-board (IOB), referring to the amount of insulin that remains active from earlier doses. This intricacy is compounded by the fact that the calculation hinges on personalized, fluctuating therapy parameters linked to the individual’s insulin sensitivity, making it a daunting challenge. Moreover, adjustments to the insulin dose may be imperative to accommodate the effects of physical activity or the influences of other macronutrients present in the meal, such as fats and proteins, which can alter glucose levels in a manner distinct from carbohydrates. (\cite{el2018artificial}). 

%These personalized parameters are the carbohydrate ratio (CR: g/U), the correction factor (CF: mg/dL/U), and a target glucose ($G_{target}$: mg/dL). A bolus can be computed with equation \ref{eq:standardbolus} given: the counted carbohydrate intake (CHO: g), the current glucose (G: mg/dL), and optionally an estimate of the insulin-on-board (IOB: U)

\begin{equation}
 B = \frac{CHO}{CR} + \frac{G - G_{target}}{CF} - IOB
 \label{eq:standardbolus}
\end{equation}

Due to these complexities, glucose control after meals remains a challenging part of diabetes and a major contributor to overall degraded glycemic control.
Still, some individuals tend to estimate their insulin doses based on their experiences with previous meals instead of performing precise CHO calculations. They depend on empirical estimates, learning through trial and error from their history of consumed meals and administered insulin doses.
%Interestingly, some people end up only estimating the  insulin dose using their experience with previously consumed meals rather than a proper calculation: they rely on empirical estimates and learn by trial and error from their previously consumed meals and delivered insulin doses.

This trial-and-error process somewhat resembles the training of a reinforcement learning (RL) algorithm.
In this work, our aim is to show that a deep-RL agent using self-attention networks can discover optimal meal boluses from an individual's historical data without any knowledge about the standard therapy parameters used in traditional bolus calculators.

\subsection{Review of RL algorithms for Bolus Calculators}

%RL is a framework for learning decision-making strategies by interacting with an environment. In this framework, an agent learns to map a situation (a state) to action, resulting in maximum cumulative future rewards \citep{sutton2018reinforcement}. 
Previous works explored RL to optimize insulin but focused primarily on adaptation to therapy parameters \citep{tejedor2020reinforcement}. Other works focused on adapting the bolus calculator but did not use RL \citep{unsworth2023adaptive}. Some have explored the use of RL to optimize bolus calculators:
(i) \citet{zhu2020insulin} proposed a method using double deep Q-learning.
The RL agent learns to select a percentage to modify the dose given by the standard bolus calculator.
This system still relies on standard bolus calculation and has a restricted action space, potentially limiting the benefits of deep learning.
%(ii) \citet{cappon2018neural} used a supervised learning approach to train a neural network to modify the standard bolus calculator.
%They use the UVA/Padova simulator and a grid search to discover the optimal bolus and train a neural network to approximate the bolus. 
(ii) \citet{ahmad2022bolus} proposes a method for automatic bolus generation without carbohydrate amounts, and only the meal type is announced (breakfast, lunch, or dinner). The authors reported problems when the meal was unexpectedly small.
% This system uses classical Q-Learning approaches and additional information may be extracted with newer deep-learning methods.
(iii) \citet{Jaloli2023} use the Soft-Actor-Critic algorithm to optimize the bolus for MDI therapy. 
Their agent learns to give a bolus based only on blood sugar and meal history, with no notion of the standard bolus calculator,
but the boluses are not user-initiated, so the system can request a bolus at any time which may increase the system burden.
(iv) \citet{el2023using} proposed a new bolus calculator based on meal categories rather than carbohydrate counting. However, the optimization algorithm needed multiple weeks to converge.

This work distinguished itself by (i) not relying on the standard therapy parameters; (ii) adjusting the insulin bolus at each call; (iii) learning directly from a two-week user's previous glucose, insulin doses, and consumed meals; (iv) using self-attention mechanisms to capture patterns.  

\subsection{Review of Proximal Policy Optimization}
Proximal Policy Optimization (PPO) is a state-of-the-art deep reinforcement learning algorithm \citep{schulman2017proximal}. 
Policy gradient-based methods (like PPO) have been shown to be very effective in high-dimensional problems with continuous action spaces. 
PPO is a \textit{on-policy} learning algorithm that makes small, constrained steps from the current policy through a clipped objective function. The loss function for PPO is given in equation \ref{eq:ppo}.

\begin{equation}
 L(\theta) = \mathbb{E}\left[ \min\left( r_t(\theta) A_t, clip(r_t(\theta), 1-\epsilon, 1+\epsilon)A_t \right) \right]
 \label{eq:ppo}
\end{equation}

where $r_t(\theta) = \frac{\pi(a_t | s_t)}{\pi_{old}(a_t | s_t)}$ is the ratio of the probabilities of taking the action of the current policy, divided by the probability of taking that action under the previous policy.
If $r_t > 1$ then the action becomes more likely under the new policy.
$A_t$ is the advantage term that defines the amount of reward (estimated via the Bellman equation) this action gives relative to the average value.

The primary loss term $r_t A_t$ intuitively means that if the advantage is positive, we want to make that action more likely, 
and if the advantage is negative, make the action less likely.
The intuition of the clipping term is that the approximation of the policy gradient is only valid near the old policy, so the loss is clipped to prevent large changes in the policy each iteration.

\section{Methods}

\begin{figure*}
 \centering
 \includegraphics[width=1\linewidth]{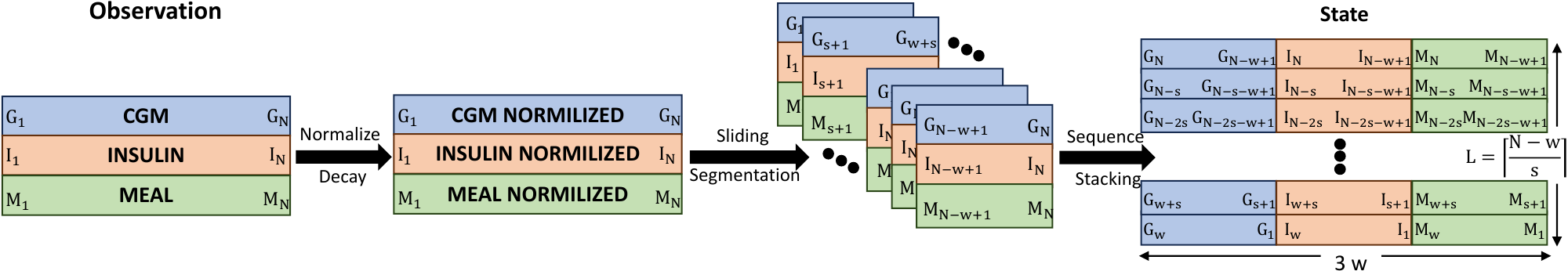}
 \caption{Steps of processing the observations: (i) data is normalized and decayed to emphasize the slow dynamics of insulin and meal, (ii) data is sliced using a window $w$ and stride $s$, (iii) data is rearranged to align periodic events.}
 \label{fig:preprocessing} 
\end{figure*}

\subsection{Problem formulation}

At a time $t_N$, we assume the availability of historical data at a fixed sampling time $t_s$: (i) measured glucose $\{\Tilde{G}_k\}_{k\in[1:N]}$, (ii) basal and bolus insulin $\{\Tilde{I}_k\}_{k\in[1:N]}$, the basal insulin is converted to amounts of insulin units delivered during the sampling time $t_s$, (iii) estimated amount of carbohydrates in meals $\{\Tilde{M}_k\}_{k\in[1:N]}$.  In the following $\mathcal{H}_k=\{\Tilde{G}_k, \Tilde{I}_k, \Tilde{M}_k\}$ is used to denote the data available at time $t_k$, and $\mathcal{H}_{1:N}$ the complete dataset.

The RL agent aims to calculate the insulin bolus $B_N$ using historical glucose levels, insulin doses, and carbohydrate estimates. Formally, it learns a function $f$ defined by:

\begin{equation} \label{eq:bolusCalc}
B_N = f(\Tilde{M}_{N}, \Tilde{G}_{N},  \mathcal{H}_{1:N-1})
\end{equation}

Note that unlike standard bolus calculation methods, which depend on predefined therapy parameters, this function leverages historical data to discern an individual's insulin sensitivity. This approach also aims to develop a universal function applicable at the population level rather than fine-tuning this function for individual nuances. Essentially, this function should learn to interpret historical data to accurately assess the upcoming meal $M_N$ and determine the optimal insulin bolus $B_N$ for any individual.

In PPO, the learning process is facilitated by an actor network and a value network. The actor network should learn to approximate $f$, and the value network should learn to approximate the \textit{value} of a state:

\begin{equation} 
V_N = g(\Tilde{M}_{N}, \Tilde{G}_{N},  \mathcal{H}_{1:N-1})
\end{equation}

%Where in this case, the history $\mathcal{H}_{1:N}$ includes $I_N = B_N + U_N$, where $B_N$ is the agent action and $U_N$ is any additional insulin delivered at $t_N$, including the delivered basal insulin.

The $f$ and $g$ are designed to: (i) identify the intrinsic delays related to the dynamics of glucose, the absorption of insulin, and the absorption of carbohydrates; (ii) convert the data into an encoded form that highlights important features for the subsequent calculation of actions or values; (iii) treat the raw data as a time series to account for recurring events that occur with regular frequency, such as eating three to four meals daily, and differentiate between night and day, as well as weekdays. To facilitate this process, we perform the following: (i) Process the raw observations to emphasize the system's dynamics as recommended in \citep{jacobs2023artificial}. (ii) Process the raw data to align periodic events. (iii) Introduce a shared function that extracts relevant features from the state: an \textit{encoder network}, before passing the encoded state to the actor and critic networks.

\subsection{Neural Network Architecture}

\subsubsection{Observations Preprocessing}

Following the international consensus on continuous glucose monitoring (CGM) metrics by \citep{battelino2019clinical}, we choose to use 14 days of historical data to calculate insulin dose. The collected data are sampled with a $t_s = 15$ minutes, resulting in a time series of length $N=1344$. 

Because of the asymmetric relevance of low glucose values compared to high glucose values, we normalize the CGM by a log-transform as shown in equation (\ref{eqn:cgmnorm}).

\begin{equation} \label{eqn:cgmnorm}
 G = \dfrac{2 \log(\Tilde{G}) - (\log(\Tilde{G}_{max}) + \log(\Tilde{G}_{min}))}{\log(\Tilde{G}_{max}) - \log(\Tilde{G}_{min})}
\end{equation}

where $\Tilde{G}_{max}=180$ and $\Tilde{G}_{min}=70$ are chosen to match hyper-/hypo-glycemic levels to 1 and -1, respectively.

We model the delays in insulin and meal absorption using the following generic decay function. 

\begin{equation} \label{eqn:fdecay}
d (k,\tau,t_s)= \left(1+\dfrac{(k-1) t_s}{\tau}\right) e^{-\dfrac{(k-1) t_s}{\tau}} 
\end{equation}

where $\tau$ is a time constant related to the time-to-maximum effects. 
The insulin and meal information is normalized using the decay function $d(.)$ and the sum of all events during the 14 days:

\begin{equation} \label{eqn:norm}
\begin{split}
I_k & = \frac{\sum_{j=1}^{k} d (k-j+1,\tau_i,t_s) \Tilde{I_j}}{\sum_{i=1}^{N} \sum_{j=1}^{i} d (i-j+1,\tau_i,t_s) \Tilde{I_j}} \\ 
M_k & = \alpha_m \frac{\sum_{j=1}^{k} d (k-j+1,\tau_m,t_s) \Tilde{M_j}}{\sum_{i=1}^{N} \sum_{j=1}^{i} d (i-j+1,\tau_m,t_s) \Tilde{M_j}} \\ 
\end{split}
\end{equation}

where $\tau_i=75$ minutes and $\tau_m=45$ minutes and $\alpha_m=\dfrac{1}{4}$ is a scaling factor that represents the inverse of the average number of meals per day.

Due to the length of the observed data, the sparsity of the insulin bolus information, and the local correlation of cgm/insulin/meals, we choose to further transform the observation, as shown in figure \ref{fig:preprocessing}. First, we perform a sliding window segmentation of the time series with width $w$ and overlap $s$. The data is then rearranged as a new sequence where each sequence element is an array of size $3 \times w$ containing related glucose/insulin/meal data. The new sequence, called state, is of length $L = \lceil \frac{N t_s - w}{s} \rceil$. We selected $w=240$ minutes and $s=120$ minutes, to achieve a reasonable shrinkage of the observation $L=166$ (grid search results of $w$ and $s$ are not shown in this work).

\subsubsection{Encoded States}

\begin{figure}
 \centering
 \includegraphics[width=1\linewidth]{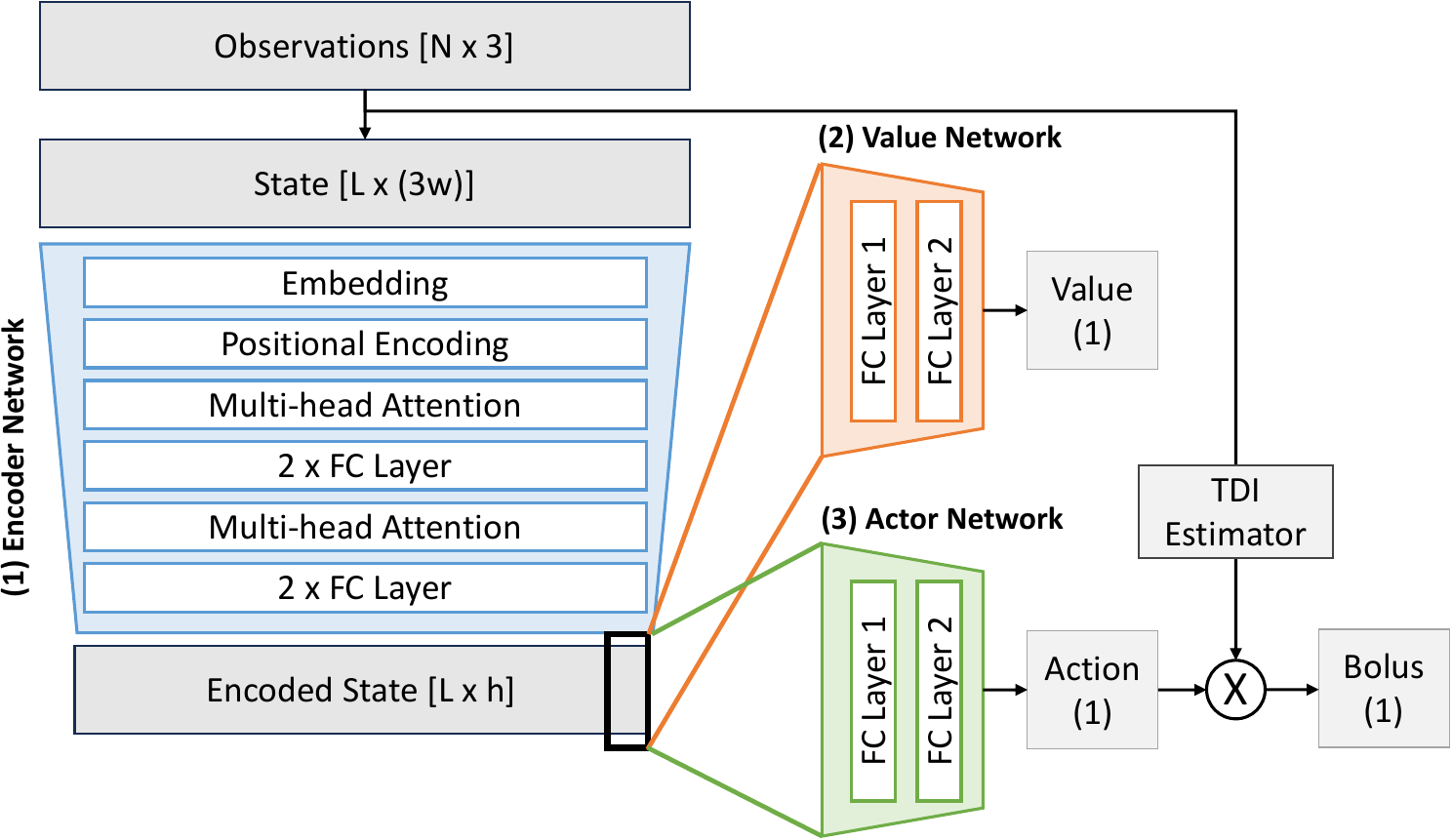}
 \caption{Neural networks architecture including (1) the encoder network (specifically the self-attention network) (2) the value network (3) the actor network.}
 \label{fig:architecture}
\end{figure}

The encoded state (output of the encoder network) is given to the actor and value network. This work aims to show that self-attention mechanisms (ATT) are key to generating the encoded state. As depicted in figure \ref{fig:architecture}, we use a standard transformer network as described in the seminal work by \citep{NIPS2017_3f5ee243}. 

Let $h$ be the size of the hidden state. In the sequence of hidden states of the form $L \times h$, only the last hidden state is used. This is done to imitate a recurrent neural network and force the last state to be the most relevant as it contains the information of the latest consumed meal and the current glucose. We chose to stack two layers of self-attention and did not explore other depths in this work.

Other encoder architectures were implemented:
\begin{itemize}
    \item LSTM: Long short-term memory network.
    \item biLSTM: Bidirectional LSTM network.
    \item FC: Fully connected layer. Instead of passing the whole state time series to the FC network only the last state of size $3 w$ is passed.
\end{itemize}
All explored architectures had 2 layers.

\subsubsection{Actions}

The action of the agent $a_N$ is defined as a fraction of the estimated total daily insulin (TDI) ($ B_N = a_N \times TDI
$). TDI is calculated directly from the observations as the total sum of insulin records per day. The action (the output of the actor-network) is bounded using a tangent hyperbolic transformation to be within 0.0 and 0.2.

\subsubsection{Rewards}

\begin{figure}
 \centering
 \includegraphics[width=1\linewidth]{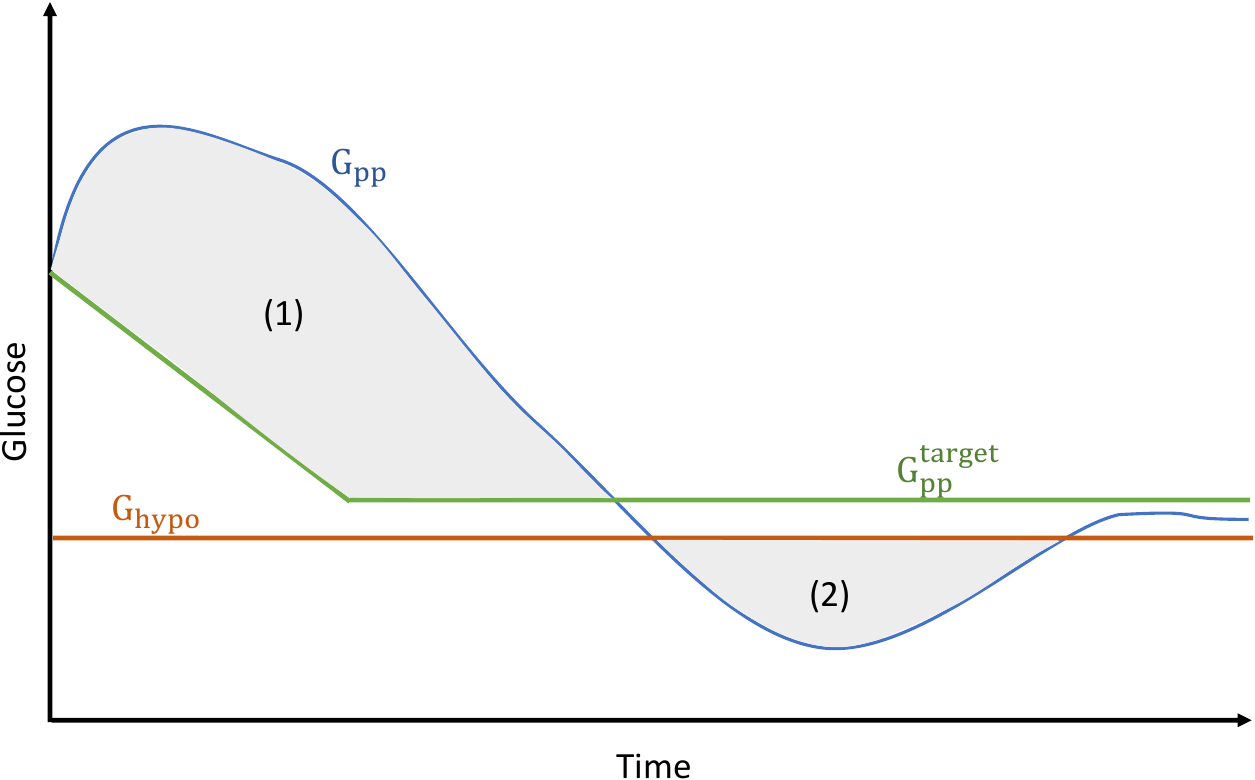}
 \caption{Depiction of reward calculation in equation \ref{eqn:reward}. The smaller the shaded grey areas (1) and (2) the bigger the reward.}
 \label{fig:reward}
\end{figure}

Following each action, the next 8 hours of glucose is extracted to estimate the reward of the action. If another action occurs within 8 hours, only the glucose before the next action is considered. We require a minimum of 3 hours of glucose data for the action to be considered for training. This signal is referred to as $G_{pp}$.

The reward $R$ is calculated using $G_{pp}$ as indicated in equation \ref{eqn:reward}.

\begin{multline} \label{eqn:reward}
 R=-\dfrac{\dfrac{1}{\#G_{pp}} \sum{\max\left(G_{pp}-G_{pp}^{target}, 0\right)} -\mu_{iAUC}}{\sigma_{iAUC}}  \\
 -\dfrac{10}{\#G_{pp}}\sum{\text{Risk}(G_{pp})^2(G_{pp}<G_{hypo})}
\end{multline}

where $G_{pp}^{target}$ is a trace connecting the glucose levels at time of the action to the desired glucose target (110 mg/dL) using a slope of $20$ mg/dL/h, $G_{hypo}=90$ mg/dL is a threshold to penalize lower glucose levels. $\#G_{pp}$ stands for the size of the $G_{pp}$ array. $\text{Risk}(G_{pp})=\log(G_{pp})^{1.084} -5.381$ is the risk function as defined by \citep{kovatchev2017metrics}. $\mu_{iAUC}=60$ mg/dL and $\sigma_{iAUC}=45$ mg/dL are tuning parameters representing the mean and standard deviation incremental area under the curve (iAUC) of glucose. Figure \ref{fig:reward} shows a visualization of the reward.

\subsection{Experimental Setup and Training}

%The training and validation scripts are implemented in Python 3.9 and networks were implemented using the pytorch 2.1 library. The agent (encoder + actor networks) is serialized using TorchScript. A C++ implementation of the FDA-accepted UVA/Padova T1D simulator is used with adults \textit{in-silico} virtual subjects (VS). Pytorch C++ API is used to call the agent within the simulator. 
The training and validation scripts were developed in Python 3.9, using PyTorch 2.1 library to implement the networks. The agent, comprising the encoder and actor networks, is serialized using TorchScript for efficiency and compatibility. We employed a C++ version of the FDA-recognized UVA/Padova T1D simulator, focusing on adult \textit{virtual subjects} (VS) for \textit{in-silico} experimentation. The PyTorch C++ API facilitated the integration of the agent within the simulator.

During the training, 80 VS were utilized. The simulations covered both sensor-augmented insulin pump (SAP) therapy, encompassing basal insulin, meal-accompanying bolus doses, and occasional correction doses for high glucose, and an AID system, specifically a legacy version of Control-IQ \citep{brown2019six}.

%Eighty VS were used during the training phase. Simulations were conducted using MDI therapy (daily basal insulin doses, meal-accompanying bolus doses, occasional high glucose correction doses) or an AID system (legacy version of control-IQ, a commercially available system).

%\footnote{\url{https://www.tandemdiabetes.com/landing-pages/t-slim-x2}}.

%Each training epoch, 20 VS were randomly selected and simulated for 21 days using both MDI and AID, and two starting random seeds (total of $20\times2\times2=80$ simulation). Seeds controlled the randomness of the experiment, including: (1) waking up time, (2) mealtimes, (3) size of consumed meals, (4) errors in CR/CF, (5) carbohydrate counting errors, (6) unannounced snacks or meals, (7) skipped meals, (8) delays between time of meal and time of insulin dose (9) errors in the optimal basal dose, (10) insulin variability during the day, (11) insulin variability in-between days. In these simulations, the agent is frozen and used to generate the insulin bolus for each announced consumed meal using a stochastic policy. The resulting 80-episode simulations are then processed to extract a sequence of transitions (observations, actions, rewards, dones) that are used in training. 

In each training epoch, 20 VS were randomly chosen for a 21-day simulation under both SAP and AID conditions, with two initial random seeds (yielding a total of $20\times2\times2=80$ simulations). These seeds introduced variability in several aspects, such as wake-up times, meal timings and sizes, errors in therapy parameters, meal announcement inaccuracies, unanticipated eating activities, meal omissions, insulin dosing delays, and interday/intraday insulin sensitivity fluctuations. The agent, whose parameters were fixed during these simulations, determined the insulin bolus for each reported meal via a stochastic policy. The resulting 80-episode simulations are then processed to extract a sequence of transitions (observations, actions, rewards, end states) that are used in training. 

A total of 30 agents were trained, using five seeds across six distinct architectural designs for the encoder network. This included two attention networks (ATT) with varying parameter counts (250K and 70K), a bi-directional LSTM, and a standard LSTM both sized similarly to the larger ATT (250K), a larger biLSTM to account for double the hidden states (500K), and a simple fully connected (FC) network with parameters comparable to the smaller ATT (70K). Key hyperparameters are presented in table \ref{tab:hyperparam}

%For validation, the remaining twenty VS were simulated for 14 days using both SAP and AID, three random seeds, and two scenarios (total of $20\times2\times3\times2=240$ simulations. Scenario 1 was similar to the training scenario with metabolic and behavioral variabilities (a challenging scenario), whereas, in scenario 2, only the insulin sensitivity variability was kept (an ideal scenario). The trained agents were compared to a baseline where a standard bolus calculator (\ref{sec:bolusCalc}) is used instead of the agent.

For validation, the remaining twenty VS underwent 14-day simulations under both SAP and AID settings, with three random seeds across two scenarios ($20\times2\times3\times2=240$ simulations). \textit{Scenario 1} replicated the training environment, introducing metabolic and behavioral variabilities, while \textit{Scenario 2} maintained only the variability in insulin sensitivity, representing an idealized condition where therapy parameters are perfectly known. The performance of the trained agents was evaluated against a standard bolus calculator to establish a baseline comparison.

%We trained $5\times6=30$ agents using five seeds and six architectures for the encoder network. Notably, two attention networks (ATT) with a small and larger number of parameters (250K and 70K). A bi-directional LSTM and a standard LSTM with the same size as the large ATT (250K). A larger biLSTM to represent the doubling of the hidden states (500K). finally a simple fully connected (FC) network with a size similar to the small ATT (70K).

\subsection{Outcomes Metrics}

Algorithm performance was assessed using the percentage time spent $<70$ mg/dL (TBR1); the percentage time spent in the glucose target $70-180$ mg/dL (TIR); and the overall glycemic risk as defined in \citep{kovatchev2017metrics}. 

\section{Results}

On average $\sim200$ patient-years were needed for training. 

Table \ref{tab:outcomes} summarizes the obtained results\footnote{Detailed results in weight and bias account: \url{https://wandb.ai/center-for-diabetes-technology/BMS24-Attention-Bolus-T1D}}. Notably ATT-based encoder networks resulted in the smallest overall glycemic risk while requiring fewer parameters. The best ATT 250K network reduced risk in all scenarios and all therapy modalities. All the trained agents outperformed the baseline in the worst-case scenario 1 while not all agents were able to match the ideal scenario 2.

\begin{table*}
\setlength\tabcolsep{3.8pt}
\begin{center}
\caption{Evaluation of six agents with different architectures and sizes on the 20 test VS. Scenario 2 represents an ideal scenario with optimal therapy parameters. Scenario 1 includes metabolic and behavior variability. The mean and standard deviation are for a 5-time repeated training. Agents with lower risk are in bold and ones with lower than the baseline are in italics.}
\label{tab:outcomes}
\begin{tabular}{l|c|c|p{0.7cm}p{0.7cm}p{0.7cm}|p{0.7cm}p{0.7cm}p{0.7cm}|p{0.7cm}p{0.7cm}p{0.7cm}|p{0.7cm}p{0.7cm}p{0.7cm}| c}
\toprule
 &  & & \multicolumn{6}{c|}{\textbf{Scenario 1}} & \multicolumn{6}{c|}{\textbf{Scenario 2}} & \textbf{Overall} \\
\cline{4-15}
& \textbf{Num} & \textbf{Hidden} & \multicolumn{3}{c|}{\textbf{SAP}} & \multicolumn{3}{c|}{\textbf{AID}} & \multicolumn{3}{c|}{\textbf{SAP}} & \multicolumn{3}{c|}{\textbf{AID}} & \textbf{Risk} \\
& \textbf{Param} & \textbf{Size} & \textbf{TIR} & \textbf{TBR} & \textbf{Risk} & \textbf{TIR} & \textbf{TBR} & \textbf{Risk} & \textbf{TIR} & \textbf{TBR} & \textbf{Risk} & \textbf{TIR} & \textbf{TBR} & \textbf{Risk} & \\
\midrule
\textbf{Baseline} & N/A & N/A & 41.5 & 3.1 & 16.5 & 59.2 & 2.0	& 9.1 & 72.2 & 2.5 & 6.2 & 77.1 & 2.1 & 5.1 & 9.2\\
\midrule
\textbf{ATT} & 250K & 96 & 58.9 (1.3) &1.5 (0.1) & \textbf{9.6} (0.3) & 69.3 (0.9) & 1.1 (0.1) &\textbf{6.7} (0.2) & 74.4 (0.2) &1.3 (0.1) & \textbf{5.7} (0.0) & 77.8 (0.5) & 1.1 (0.1) &\textbf{5.0} (0.1) &\textbf{6.8} \\
\textbf{ATT} & 70K & 48 & 57.3 (1.0) &1.4 (0.0) &\textit{10.0} (0.3) & 69.4 (0.5) & 1.1 (0.0) &\textbf{6.7} (0.1) & 73.4 (0.2) &1.2 (0.1) & \textit{5.9} (0.0) & 77.7 (0.1) & 1.1 (0.0) &\textit{5.1} (0.0) &\textit{6.9} \\
\textbf{biLSTM} & 500K & 192 & 53.8 (1.0) &1.3 (0.1) &\textit{10.9} (0.2) & 68.8 (1.0) & 1.1 (0.1) &\textit{6.8} (0.2) & 72.7 (1.0) &1.0 (0.1) & \textit{6.1} (0.2) & 77.6 (1.0) & 1.2 (0.1) &\textbf{5.0} (0.2) &\textit{7.2} \\
\textbf{biLSTM} & 250K & 128 &  53.2 (0.8) &1.1 (0.1) &\textit{11.0} (0.2) & 67.4 (0.3) & 1.0 (0.1) &\textit{7.1} (0.0) & 70.0 (0.2) &0.8 (0.1) & 6.5 (0.0) & 75.4 (0.2) & 1.0 (0.1) &5.4 (0.0) & \textit{7.5} \\
\textbf{LSTM} & 250K& 128 & 53.6 (0.9) &1.1 (0.1) &\textit{10.8} (0.3) & 67.9 (0.1) & 1.0 (0.1) &\textit{7.0} (0.0) & 70.9 (0.3) &0.9 (0.1) & 6.3 (0.1) & 76.3 (0.4) & 1.1 (0.1) &5.3 (0.1) &\textit{7.4} \\
\textbf{FC} & 70K & 128 &  47.0 (1.3) &1.2 (0.2) &\textit{12.8} (0.3) & 65.5 (0.8) & 1.1 (0.1) &\textit{7.4} (0.1) & 65.2 (1.7) &0.9 (0.1) & 7.3 (0.3) & 75.4 (1.0) & 1.1 (0.1) &5.4 (0.2) & \textit{8.2} \\
\bottomrule
\end{tabular}
\end{center}
\end{table*}

\begin{figure*}
 \centering
 \includegraphics[width=1\linewidth]{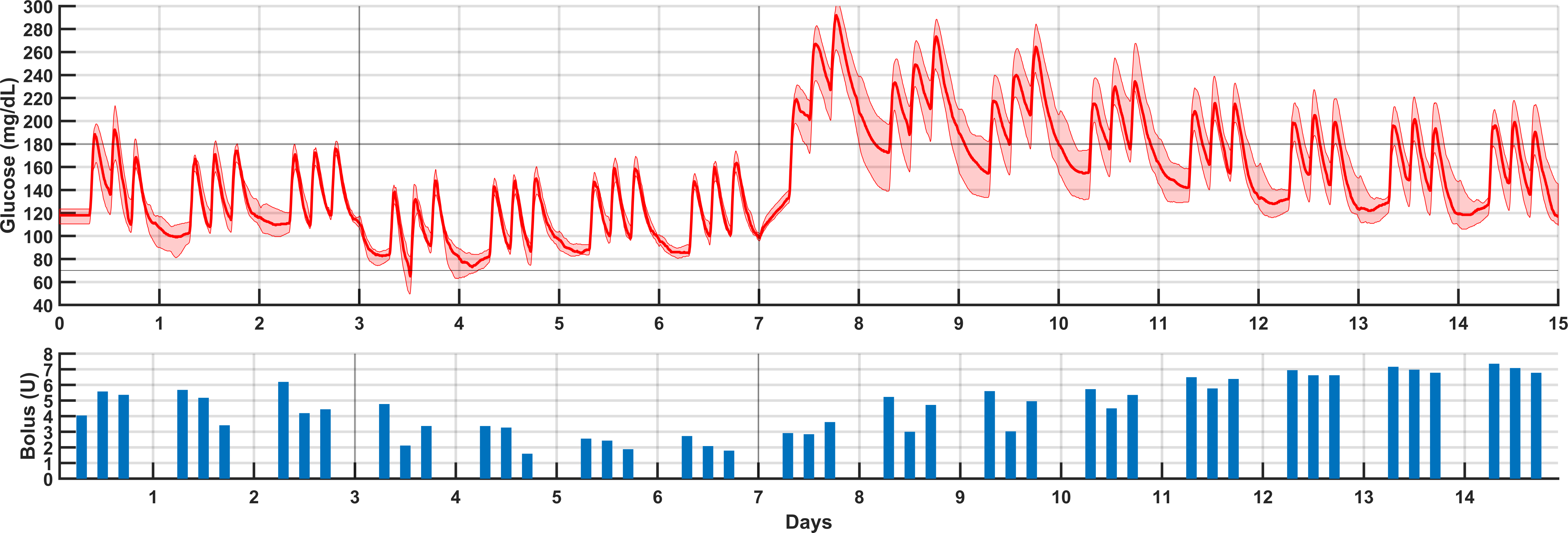}
 \caption{Test experiment of 20 VS. The same meal (60g) is given 3-times a day. Insulin sensitivity is doubled on day 3 and halved on day 7.}
 \label{fig:test}
\end{figure*}

Figure \ref{fig:test} presents the results of a robustness experiment where the ATT 250K agent is used in a simulation with the twenty validation VS. VS consumed the same meal (60g) three times per day and insulin sensitivity was abruptly doubled at day 3 then abruptly halved at day 7. While the agent has never been explicitly trained in such a scenario it was able to modulate the insulin doses quickly and safely to recover from the abrupt changes in insulin sensitivity.

\section{Discussion}

%We explored the \textit{in-silico} feasibility of an RL agent that leverages attention mechanisms to optimize meal-accompanying insulin doses without the need for standard therapy parameters (CR/CF) but using historical glucose, insulin, and meal information. This RL agent is expected to operate on a smartphone that can connect via Bluetooth to diabetes management devices, including insulin pumps, smart insulin pens, and continuous glucose monitoring systems. 

We investigated the \textit{in-silico} feasibility of an RL agent that employs attention mechanisms to refine insulin dosing at mealtimes, circumventing the need for traditional therapy parameters (CR/CF) by utilizing past data on glucose levels, insulin doses, and meal information. This RL agent is envisioned to be at the core of a smartphone app that communicates with diabetes management devices like insulin pumps, smart insulin pens, and CGM.

%The results in table \ref{tab:outcomes} showcase the ability of attention networks to efficiently encode historical data compared to other architectures. Even with a small hidden state of size $3 \times w=48$ and the small number of parameters $70K$, the self-attention encoder results in competitive outcomes.

Table \ref{tab:outcomes} demonstrates the superior efficiency of attention networks in processing historical data over other architectures. Notably, a self-attention encoder achieved competitive performance with a modest hidden state size of $3 \times w=48$ and a parameter count of $70K$. Furthermore, the ATT 250K model notably outperformed the baseline in Scenario 2, a challenging feat given that Scenario 2 employs theoretically optimal therapy parameters optimized for single-meal instances. The experiment in figure \ref{fig:test} also reveals that the agent adjusted differently the insulin doses for different times of the day, enhancing efficacy beyond the single-meal optimization paradigm. Specifically, the initial meal typically received a larger insulin dose relative to subsequent meals.

%Results also show how the ATT 250K was able to even surpass the baseline in Scenario 2. Theoretically, scenario 2 uses theoretically optimal therapy parameters which makes it difficult to improve. However, these therapy parameters were optimized for a one-meal scenario. By looking at the experiment \ref{fig:test} we see that the agent further optimized the insulin dose for consecutive meals in a way that surpassed what could have been don in a one-meal scenario. Essentially, the first meal of the day seems to receive a bigger dose compared to the 2 others.

%The way we formulated the reward function is one of the novelties of this work. The most common approach is to use a derivative of glucose risk \citet{zhu2020basal}. In equation \ref{eqn:reward} the use of $G^{target}_{pp}$ ensures to not inadequately punish actions happening at high glucose levels but rather punish the time needed to achieve the target. 

A novel aspect of our research is the formulation of the reward function combining a hypoglycemia term derived from the glucose risk and a hyperglycemia term derived from iAUC. As equation \ref{eqn:reward} illustrates, we focus on achieving a target postprandial glucose level ($G^{target}_{pp}$) without unduly penalizing actions at high glucose levels, instead emphasizing the duration to reach the target. This work's other distinctive strength lies in its training across two therapeutic approaches—SAP and AID, showing the agent's effectiveness in both contexts. This also hints that the agent might have learned to take into account how basal insulin is delivered. 

Another noteworthy aspect is the agent's ability to adapt to different sizes of observations, despite being optimized to handle observation periods of two weeks. This can be attributed to the way tokens were constructed (concatenated 4-hour CGM/Insulin/Meal) (Fig. \ref{fig:preprocessing}). This adaptability was further evidenced in (Fig. \ref{fig:test}) where the agent successfully functioned after being fed only eight hours of nighttime data as seen in the first meal of the experiment.

%A strong characteristic of this work is that it was trained in two different therapy modalities: SAP and AID and the agent is as good in both. Because of the way it was trained we ensured that the insulin signal was taken into consideration. Previous work (\cite{el2023using}) only used meal categorization instead of carbohydrate counting but did not consider the local dynamics of each bolus individually. A combination of these works would be interesting as future work: simplifying the carbohydrate counting process while leveraging local information for individual bolus. 

%Results in figure \ref{fig:test} also showed that after halving the sensitivity the agent was struggling fo further increase the insulin dose following day 13. This is due to a limitation we imposed on each dose to be under 20\% of the estimated TDI, such a limitation will be reviewed in further work.  Other limitations include the reliance on the simulation environment and only exploring six different architectures

Figure \ref{fig:test} also highlighted a challenge encountered by the agent in increasing insulin doses after a reduction in sensitivity beyond day 13, attributed to a self-imposed cap of 20\% of the estimated Total Daily Insulin (TDI). This limitation, among others like the lack of use of real-world data dependency and the exploration of only six architectures can be addressed in future work to enhance the model's applicability and accuracy. Future research could also integrate our findings with methods that simplify carbohydrate counting (\cite{el2023using}), utilizing local bolus information to further refine dosing strategies.

 \section{Conclusion}

We introduce an RL agent designed to refine insulin dosing strategies at mealtimes for individuals with T1D, eliminating the necessity for personalized therapy parameters. The utilization of self-attention mechanisms was crucial for effectively processing individual historical data. This innovative approach promises to reduce the dependency on optimizing therapy parameters, thereby streamlining the treatment process and potentially enhancing glycemic control. These results remain preliminary and future research aims to further simplify the process of carbohydrate counting and incorporate real-world data to validate and extend our findings.

%An RL approach to optimize insulin doses in a novel mealtime dosing paradigm not requiring personalized therapy parameters for people with T1D is presented. Self-attention was key to leveraging the individual historical data. This new paradigm has the potential to alleviate the need for therapy parameters optimization, thus simplifying the therapy and improving glycemic outcomes. These results remain preliminary and limited to the conducted \textit{in-silico} experiments. Further research will explore the simplification of carbohydrate counting and the use of real-world data.

% \begin{ack}
% Place acknowledgments here.
% \end{ack}

\bibliography{ifacconf} % bib file to produce the bibliography
 % with bibtex (preferred)

\appendix
\section{Hyperparamters}

\begin{table}[H]
\begin{center}
\caption{Key training hyper-parameters}\label{tab:hyperparam}
\begin{tabular}{cc|cc}
\toprule
\textbf{Parameter} & \textbf{Value} & \textbf{Parameter} & \textbf{Value} \\
\midrule
%Maximum number of epochs & $50$ \\
Batch Size & $1024$ & Discount factor $\gamma$ & 0.10 \\
Learning rate & 0.001 & PPO entropy coefficient & 0.1\\
Dropout & 0.1 & PPO Target KL-divergence & 0.05 \\
GAE lambda & 0.99 & PPO Clipping coefficient & 0.05 \\
% \\
%Depth of encoder network & $2$ \\
%Depth of actor network & $2$ \\
%Depth of critic network & $2$ \\
\bottomrule
\end{tabular}
\end{center}
\end{table}

\end{document}